## Review Articcle

# First Record of *Echinoparyphium recurvatum* (Trematoda, Echinostomatidae) in Pakistan with New Avian Definitive Host, *Vanellus leucurus*

**Nadir Ali Birmani[1], Saima Naz,[1] Gabriel Mouahid[2*]**

[1]Department of Zoology, University of Sindh, Jamshoro, Sindh, Pakistan

[2]IHPE UMR 5244 CNRS/UPVD/IFREMER/UM/ University of Perpignan, Perpignan, France

***Corresponding author:** Gabriel Mouahid, IHPE UMR 5244 CNRS/UPVD/IFREMER/UM/ University of Perpignan, 58 avenue Paul Alduy, 66860 Perpignan, France. Tel: +33-468662184; Fax: +33-468662281; Email: mouahid@univ-perp.fr

## Abstract

Parasitological examinations of White-tailed Lapwing *Vanellus leucurus* (Charadriiformes: Charadriidae) collected from Dadu district (Sindh Province, Southern Pakistan) were carried out at the Department of Zoology, University of Sindh, Jamshoro. A total of 64 trematodes belonging to the genus *Echinoparyphium* [1], were collected from the small intestine of two individual hosts. On the basis of the number and varying size of collar spines, number and size of ventral spines, body shape, arrangement of genital organs, distribution of vitellaria and other diagnostic characters, these trematodes were identified as *Echinoparyphium recurvatum* [2]. Our study provides the first Pakistan record of the trematode *E. recurvatum* with also the first record of *V. leucurus* as avian definitive host.

## Introduction

Genus *Echinoparyphium* is an important taxon in the family Echinostomatidae and has considerable importance in medical and veterinary sciences [3]. It has previously been described as *Distomum* by von Linstow [2] using specimens collected from the small intestine of *Fuligula manila*. Dietz [1] carried out a revision of Echinostomatidae and proposed different genus including the genus *Echinoparyphium*. Kanev [4] published a checklist with 151 species of *Echinoparyphium* using freshwater snails as first and second intermediate hosts, tadpoles and fish as a second intermediate hosts, and birds and mammals as definitive hosts. Since the creation of the genus, most of the published reports are based on the adequate descriptions of adult specimens. The more differentiating character is the morphometry of collar spines in adult stage, specifically the total number of spines (29-45), their sharply pointed shape and their arrangement in double row, with dorsal aboral spines noticeably longer than dorsal oral ones.

In the present paper, we report for the first time the presence of *E. recurvatum* recovered from the small intestine of the White-tailed Lapwing *Vanellus leucurus* (Charadriiformes: Charadriidae) in the Sindh province (Southern Pakistan) and describe the morphology of the adult. The White-tailed Lapwing is inhabitant of freshwater marshy areas and lake shores. They more frequently seek food in shallow water. They are omnivorous and their food includes insects (chiefly beetles, grasshoppers, caterpillars,and fly larvae), mollusks, worms and crustaceans [5] which are potential sources of developmental stages of a variety of helminth parasites.

## Materials And Methods

Four live White-tailed Lapwings *Vanellus leucurus* were collected from the Dadu city (GPS coordinates: 26°44'2.76"N; 67°46'46.20"E) located in the North-Western Sindh province, Pakistan. Birds were examined for the presence of trematode parasites in the Parasitology laboratory of the department of Zoology, University of Sindh, Jamshoro, in January 2016. A total of 64 trematodes belonging to the genus *Echinoparyphium*, were collected from the small intestine of two birds and were processed according to the method given by Garcia and Ash [6]. Line drawings were made with the help of a camera Lucida. All measurements were based on 13 mature specimens and are given in micrometer (µm) followed by range in parenthesis. For practical reasons, we have chosen to give the value of the minor axis and the major axis for any spherical or subspherical organ. Length and width are kept for elongated structures such as the body or its various parts, oesophagus, head collar spines, etc. Identification was made with the help of keys given by Jones *et al.* [7] andYamaguti [8]. Specimens are available at the Museum of Natural History in Paris under the voucher reference: MNHN HEL682.

## Taxonomic Part

**Description** (based on 13 mature specimens; (Figures 1-3; Table 1)

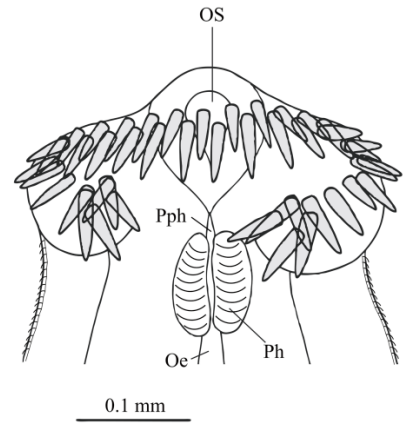

**Figure 2:** Morphology of *Echinoparyphium recurvatum*: anterior end of the adult with collar spined arrangement (45 spines), Oral Sucker (OS), Prepharynx (Pph), Pharynx (Ph) and Oesophagus (Oe).

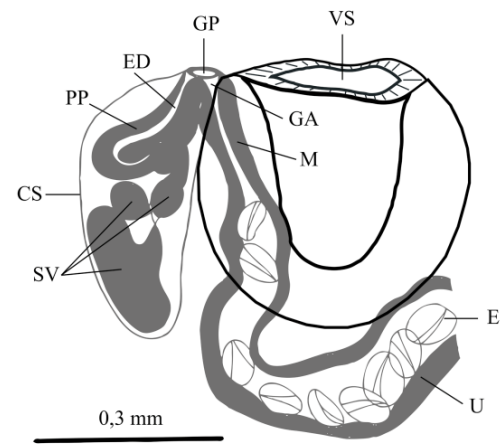

**Figure 3:** Morphology of *Echinoparyphium recurvatum*: genital pore region with Ventral Sucker (VS), Uterus (U) with Eggs (E), end part of the uterus or Metraterm (M), Genital Atrium (GA) and cirrus sac with Seminal Vesicle (SV), Pars Prostatica (PP), Ejaculatory Duct (ED) and Genital Pore (GP).

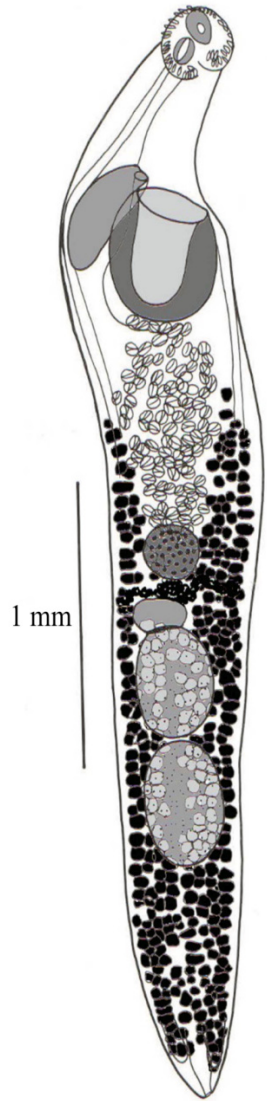

**Figure 1:** Morphology of *Echinoparyphium recurvatum*: Lateroventral view of entire adult worm.

Body small, spinose, elongate, tapering posteriorly, 3868 (3150-3905) length by 684 (521-690) width. Maximum width at level of ventral sucker. Forebody small, measuring 657 (600-682) length ventrally bent slightly. Hindbody measuring 2736 (2051-2818) length. Integument armed with very small spines which extend from the head crown to the testicular region. Head collar reniform, well developed, measuring 305 (288-310) by 263 (220-

271) and armed with 45 spines arranged in double rows. Corner oral spines 6 on each side, 53.5 (35-70) length by 15.5 (9-22) width; lateral spines 39 (30-46) length by 10 (9-13) width, dorsal spines on oral sucker region 75.5 (40-90) length by 12.5 (9-14) width; dorsal spines away from oral sucker 98.5 (80-120) length by 39 (30-50) width. Oral sucker subterminal, pear shaped in most of the specimens, measuring 118 (110-125) by 110 (100-118). Pre-pharynx very small, 14.5 (14-18) length. Pharynx muscular, elongate-oval, measuring 110 (106-120) by 78 (70-80). Oesophagus small, tubular, measuring 394 (368-422) length. Intestinal bifurcation in front of ventral sucker. Ventral sucker large, muscular, cup-shaped, measuring 473 (425-500) by 421 (392-430). Testes tandem, elongate-oval, smooth, contiguous, in some specimens slightly lobed, located in posterior half of the body. Anterior testicle measuring 394 (385-405) by 315 (302-320) located in the 3rd quarter of the body, whereas, a major part of posterior testicle present in the 3rd quarter and very small portion of it extend to last 4th quarter of body. Posterior testicle measuring 447 (430-455) by 315 (310-320). Post-testicular space measuring 815 (810-820) length. Cirrus sac muscular, elongate, located at antero-lateral side of the ventral sucker, measuring 440 (410-450) by 196 (180-210). Genital pore preacetabular in position, funnel like in appearance. Ovary subspherical, median, measuring 223 (220-235) by 184 (180-206). Distance between ventral sucker and ovary, 789 (700-800) length. Mehlis gland distinct. Uterine seminal receptacle immediately between anterior testicle and the acetabulum, laterally elongate, measuring 210 (200-220) by 105 (100-110). Uterus without loops, filled with eggs, covering an area between ovary and ventral sucker. Eggs operculated, measuring 79 (60-98) length by 51 (35-65) width. Vitellaria composed of large, irregular follicles, distributed in lateral fields of the body, commencing from mid-level of the uterus, extending backward reaching up to posterior extremity, confluent in post-testicular area in few specimens. Excretory vesicle tubular. Excretory pore terminal.

## Discussion

The genus *Echinoparyphium* is distinct from other genera of Echinostomatidae in possessing a well-developed head collar with 29 to 45 spines in a double row, with small spines covering the tegument extending to the testicular level ventrally and to the anterior margin of the ventral sucker dorsally. Large testes are tandem, elongated-oval, smooth or irregular, contiguous, post-equatorial and located in center of the hind body. Ovary small, spherical, median or submedian, pre-equatorial [9]. The genus *Echinoparyphium* counts many parasitic species of the intestine of many groups of birds and some species of mammals and the Type species is *E. elegans* [3,10,11]. Several species of genus *Echinoparyphium* have been found to infect the gastrointestinal tract of aquatic birds and mammals worldwide [12,13]. They are known to infect a large number of different taxa of birds (Anseriformes, Charadriiformes, Ciconiiformes, Columbiformes, Falconiformes, Galliformes, Gruiformes, Passeriformes, Podicipediformes, Strigiformes) and few taxa of mammals (Muridae (*Hydromys chrysogaster*, *Rattus norvegicus*, *Rattus colletti* and *Rattus rattus*), Cricetidae (*Ondatra zibethicus*), Canidae (*Alopex lagopus*) and Hominidae (*Homo sapiens*) [3,14].

According to the morphological description (Figure 1-3) and to the measurements (Table 1), our specimens belong to the genus *Echinoparyphium* and to the species *recurvatum*. It is the first time that this species is described in Pakistan and the first time that the avian host White-tailed Lapwing *Vanellus leucurus* is found to be a definitive host. Khan *et al.* [15] reported two species of *Echinoparyphium* in Pakistan: (i) *Echinoparyphium ellisi* [16] from *Anas crecca* in Nowshera (Khyber Pakhtunkhwa, one of the four provinces of Pakistan located in the northwestern Region) which belongs to the "recurvatum" group according to the phylogenetic study of Kostadinova *et al.* [17] and differs from our specimens in having a very long body (6200 x 1048 µm compared to our specimens 3868 x 684µm); (ii) *Echinoparyphium recurvatum* from *Anas crecca* in Peshawar (Khyber Pakhtunkhwa) which differs from our specimens in having 37 collar spines instead of 45 as it is well known for this species. The species known to have 37 collar spines is *E. serratum* described by Howell [18] and not *E. recurvatum*.

An updated list of natural avian hosts of *E. recurvatum* along with geographical distribution is given (Table 2). It shows that *E. recurvatum* is cosmopolitan species with a wide geographical distribution associated to a wide range of avian definitive host taxa including 7 orders and 10 families. The lack of specificity towards the definitive hosts is assigned to a low specificity towards the first and second intermediate hosts [19].

| Characters | Lee *et al.*[20] | Sohn [21] | Sereno-Uribe *et al.*[9] | Present study |
|---|---|---|---|---|
| Body (L) | 3500-4700 | 2010-3090 (2760) | 2750-3220 (2940) | 3150-3905 (3868) |
| Body (W) | 500-650 | 460-610 (550) | 400-550 (460) | 521-690 (684) |
| Head collar (Ma) | 290-340 | 260-316 (283) | 190-300 (269) | 288-310 (305) |
| Head collar (Mi) | ng | 173-214 (188) | 170-220 (200) | 220-271 (263) |
| Corner oral spines (L) | ng | 53-70 | 49-66 (57) | 39-68 (53) |
| Corner oral spines (W) | ng | ng | 9-12 (10) | 11-20 (16) |
| Dorsal oral spines (L) | ng | ng | 45-62 (54) | 58-93 (65) |
| Dorsal oral spines (W) | ng | ng | 6-12 (9) | 11-14 (12) |
| Oral sucker (Ma) | 120-150 | 92-122 (109) | 78-120 (96) | 110-125 (118) |
| Oral sucker (Mi) | 120-150 | 92-112 (99) | 93-110 (100) | 100-118 (110) |
| Pharynx (L) | 110-130 | 61-87 (77) | 100-108 (90) | 106-120 (110) |
| Pharynx (W) | 90-110 | 36-71 (46) | 50-60 (60) | 70-80 (78) |
| Oesophagus (L) | 160-190 | 204-377 (315) | 270-430 (350) | 368-422 (394) |
| Cirrus sac (L) | 300-400 | 209-337 (263) | 190-260 (230) | 410-450 (440) |
| Cirrus sac (W) | 130-180 | 82-143 (117) | 90-150 (150) | 180-210 (196) |
| Ventral sucker (Ma) | 320-400 | 306-383 (337) | 260-380 (320) | 425-500 (473) |
| Ventral sucker (Mi) | 320-390 | 296-367 (333) | 240-320 (290) | 392-430 (421) |
| Ovary (Ma) | 130-220 | 92-153 (122) | 110-140 (120) | 220-235 (223) |
| Ovary (Mi) | 130-200 | 82-133 (112) | 100-130 (120) | 180-206 (184) |
| Anterior testis (L) | 280-420 | 235-377 (329) | 200-360 (300) | 385-405 (394) |
| Anterior testis (W) | 210-280 | 153-214 (188) | 90-220 (150) | 302-320 (315) |
| Posterior testis (L) | 390-480 | 265-408 (355) | 280-450 (370) | 430-455 (447) |
| Posterior testis (W) | 180-290 | 163-204 (187) | 110-180 (140) | 310-320 (315) |
| Egg (L) | 82-97 | 96-105 (102) | 70-96 (85) | 65-93 (79) |
| Egg (W) | 54-59 | 64-71 (68) | 40-56 (47) | 37-65 (51) |

**Table 1:** Comparative morphometrics (in μm) of adult worms of *Echinoparyphium recurvatum* obtained in different studies; length (L), width (W), major axis (Ma), minor axis (Mi), ng: not given and mean values given in parentheses after ranges.

| Host birds | Order and family | Geographical distribution | Reference |
|---|---|---|---|
| *Phasianus colchicus* | Galliformes, Phasianidae | Texas, USA | Pence *et al.* [22] |
| *Meleagris gallopavo* | Galliformes, Phasianidae | Eastern Kansas, USA | McJunkin *et al.* [23] |
| *Anas clypeata, Gallus gallus* | Anseriformes, Anatidae, Galliformes, Phasianidae | Mexico | Sereno-Uribe *et al.* [9] |

| *Gallinula chloropus galeata, Phimosus infuscatus* | Gruiformes, Rallidae, Pelecaniformes, Threskiornithidae | Buenos Aires Province Argentina | Lunaschi *et al*. [24] |
|---|---|---|---|
| *Charadrius hiaticula* | Charadriiformes, Charadriidae | Slovak Republic | Macko *et al.* [25] |
| *Buteo buteo* | Falconiformes, Accipitridae | Czech Republic | Sitko [26] |
| *Anser anser* | Anseriformes, Anatidae | Gilan province, Iran | Hosseini *et al.* [27] |
| *Gallinago gallinago, Tadorna ferruginea* | Charadriiformes, Scolopacidae, Anseriformes, Anatidae | India | Mehra [28] |
| *Vanellus leucurus* | Charadriiformes, Charadriidae | Sindh Province, Pakistan | Current paper |
| *Larus dominicanus* | Charadriiformes, Laridae | Otago Peninsula New Zealand | Latham and Poulin [19] |
| *Chroicocephalus novaeseelandia* | Charadriiformes, Laridae | Otago Peninsula New Zealand | Fredensborg *et al.* [29] |
| *Aythya novaeseelandia* | Anseriformes, Anatida | Lake Wanaka and the Waitaki River watershed, New Zealand | Davis [30] |
| *Anas spp., Anser anser, Aythya novaeseelandiae, Botaurus poiciloptilus, Branta canadensis, Carduelis chloris, Cygnus atratus, Larus dominicanus, Tadorna variegata,* | Anseriformes, Anatidae, Anseriformes, Anatidae, Anseriformes , Anatidae, Pelecaniformes, Ardeidae, Anseriformes, Anatidae, Passeriformes, Fringillidae, Anseriformes, Anatidae, Charadriiformes, Laridae, Anseriformes, Anatidae | New Zealand | McKenna [31] |

**Table 2:** Natural avian hosts of E*chinoparyphium recurvatum* and their geographical location.